# Giant Magnetoelectric Effect via Strain-Induced Spin-Reorientation Transitions in Ferromagnetic Films

N. A. Pertsev

*A. F. Ioffe Physico-Technical Institute, Russian Academy of Sciences, 194021 St. Petersburg, Russia*

(arXiv:cond-mat, 28 April 2008)

It is shown theoretically that a giant magnetoelectric susceptibility exceeding $10^{-6}$ s/m may be achieved in the ferromagnetic/ferroelectric epitaxial systems via the magnetization rotation induced by an electric field applied to the substrate. The predicted magnetoelectric anomaly results from the strain-driven spin-reorientation transitions in ferromagnetic films, which take place at experimentally accessible misfit strains in $CoFe_2O_4$ and Ni films.

The magnetoelectric effect, defined broadly as a coupling between magnetic and electric properties of a material system, currently attracts great interest of physicists [1-3]. In particular, the switching of magnetization direction by an electric field represents the subject of cutting-edge research [4-7] owing to potential applications in novel magnetoelectric devices such as electric-write magnetic-read memories [8]. Since the direct magnetoelectric coupling is usually weak [2], the switching caused by the strain-mediated indirect coupling between electric field and magnetization becomes an attractive possibility. This scenario may be realized via the strain-induced spin-reorientation transitions in ferromagnetic films.

Spin-reorientation transitions (SRTs) were observed in thin films of many ferromagnetic substances [9-15]. These transitions represent a change of magnetization orientation, which occurs when the film thickness exceeds a critical value. The critical film thickness usually corresponds to a coverage of several monolayers [9,13], but may be as large as a few hundreds of nanometers [15]. The size-driven SRTs in ultrathin films are explained by a thickness-dependent contribution to the film free energy, which results from the surface magnetocrystalline anisotropy [16]. Another important contribution is due to the magnetoelastic energy associated with the coupling between magnetization and lattice strains [17,18]. Since epitaxial thin films are usually strained by a dissimilar thick substrate, in some ferromagnetic films the magnetization reorientation is caused by the relaxation of lattice strains with increasing film thickness, which results from the generation of misfit dislocations [11,13,15]. This experimental observation indicates that the magnetization reorientation can be induced by tuning the film lattice strains externally. Such strain-induced transitions may be realized via the substrate bending or by the application of electric field to a piezoelectric substrate. The second method leads to the sought magnetoelectric switching.

In this paper, the strain-driven SRTs are described theoretically using a nonlinear thermodynamic approach developed earlier [19-21]. The calculations are performed for relatively thick (001)-oriented films of cubic ferromagnets, where the surface effect on the magnetization orientation may be neglected. It is shown that in $CoFe_2O_4$ and Ni films the magnetization reorientation occurs at experimentally accessible critical strains of small magnitude. For the ferromagnetic/ferroelectric heterostructures, where the substrate has a high piezoelectric response, a giant magnetoelectric susceptibility is predicted.

Consider a thin ferromagnetic film grown on a dissimilar thick nonmagnetic substrate. Since there are no mechanical forces acting on the upper film surface, the Helmholtz free energy may be used to determine its equilibrium thermodynamic state. For cubic ferromagnets, the contribution to the Helmholtz free energy density, which is associated with the energy of magnetocrystalline anisotropy, can be written as [19]

$$U_{an} = K_1(m_1^2 m_2^2 + m_1^2 m_3^2 + m_2^2 m_3^2) + K_2 m_1^2 m_2^2 m_3^2, \quad (1)$$

where $K_1$ and $K_2$ are the anisotropy constants of fourth and sixth order at constant strains, and $m_i$ ($i = 1,2,3$) are the direction cosines of the spontaneous magnetization $\mathbf{M}_s$ relative to the principal cubic axes. The elastic energy contribution equals

$$U_{el} = \frac{1}{2} c_{11}(u_{11}^2 + u_{22}^2 + u_{33}^2) + c_{12}(u_{11}u_{22} + u_{11}u_{33} + u_{22}u_{33}) + \frac{1}{2} c_{44}(u_{12}^2 + u_{13}^2 + u_{23}^2), \quad (2)$$

where $c_{11}$, $c_{12}$, and $c_{44}$ are the elastic stiffnesses at constant magnetization, and $u_{ij}$ ($i, j = 1,2,3$) are the lattice strains defined in the crystallographic reference frame ($x_1$, $x_2$, $x_3$) with the $x_3$ axis orthogonal to the film surfaces. Owing to the coupling between the magnetization and lattice strains, a magnetoelastic contribution $U_{me}$ also exists [17,18], which is usually approximated by the formula

$$U_{me} = B_1 \left[ \left( m_1^2 - \frac{1}{3} \right) u_{11} + \left( m_2^2 - \frac{1}{3} \right) u_{22} + \left( m_3^2 - \frac{1}{3} \right) u_{33} \right] + B_2 \left( m_1 m_2 u_{12} + m_1 m_3 u_{13} + m_2 m_3 u_{13} \right) \quad (3)$$

involving two magnetoelastic coefficients, $B_1$ and $B_2$. The lattice strains $u_{ij}$ in Eqs. (2) and (3) can be calculated via mechanical boundary conditions of the problem using the magnetoelastic equations of state $\sigma_{ij} = \partial(U_{el} + U_{me})/\partial u_{ij}$, where $\sigma_{ij}$ are the mechanical stresses. When the ferromagnetic layer is in a single-domain state and there are no misfit dislocations in the film/substrate system, the strain and stress fields inside the film may be assumed to be homogeneous. In this case, the lattice matching at the interface gives $u_{11} = u_{m1}$, $u_{22} = u_{m2}$, and $u_{12} = 0$, where $u_{m1} = (b_1 - a)/a$ and $u_{m2} = (b_2 - a)/a$ are the misfit strains defined by the differences between the substrate lattice parameters $b_1$ and $b_2$ measured in the directions parallel to the film in-plane crystallographic axes and the lattice constant $a$ of a free-standing film [22]. From the conditions $\sigma_{13} = \sigma_{23} = \sigma_{33} = 0$, which follow from the absence of tractions acting on the film surface, one further obtains $u_{13} = -B_2 m_1 m_3 / c_{44}$, $u_{23} = -B_2 m_2 m_3 / c_{44}$, and $u_{33} = -[B_1(m_3^2 - 1/3) + c_{12}(u_{m1} + u_{m2})]/c_{11}$. The substitution of calculated strains into Eqs. (2)-(3) and the summation of all contributions including the magnetostatic energy yields the following expression for the energy density $\Delta F$ of a homogeneously magnetized film:

$$\Delta F = B_1(u_{m1} m_1^2 + u_{m2} m_2^2) + \left[ \frac{1}{2} \mu_0 M_s^2 - \frac{B_1^2}{6c_{11}} - \frac{c_{12}}{c_{11}} B_1(u_{m1} + u_{m2}) \right] m_3^2 + K_1 m_1^2 m_2^2 + \left( K_1 + \frac{B_1^2}{2c_{11}} - \frac{B_2^2}{2c_{44}} \right) (m_1^2 + m_2^2) m_3^2 + K_2 m_1^2 m_2^2 m_3^2, \quad (4)$$

where terms independent of the magnetization direction were omitted. The single-domain state is assumed here in order to evaluate the remnant magnetization appearing after the application of a strong magnetic field (this magnetization is used to determine SRTs experimentally). It should be noted that the magnetocrystalline coefficient $K_1$ involved in Eq. (4) differs from the bulk anisotropy constant $K_{1\sigma}$, which is measured experimentally at constant stresses, but can be easily calculated as

$K_1 = K_{1\sigma} + B_1^2 c_{11}/[(c_{11} - c_{12})(c_{11} + 2c_{12})] + B_2^2/(2c_{44})$.

Equation (4) also represents a good approximation for the mean energy density of a thick film with large densities $\rho_1$ and $\rho_2$ of misfit dislocations at the film/substrate interface. Indeed, when the dislocation spacing is much smaller that the film thickness, the influence of dislocation arrays on the mean lattice strains can be described by replacing the substrate lattice parameters $b_1$ and $b_2$ in the relations for $u_{m1}$ and $u_{m2}$ by effective parameters $b_1^* = b_1(1-\rho_1)$ and $b_2^* = b_2(1-\rho_2)$ [23].

The most remarkable feature of Eq. (4) is the presence of terms linearly dependent on the misfit strains $u_{m1}$ and $u_{m2}$. Since these strains may be both positive and negative, the mechanical substrate effect may change the direction of the easy axis of magnetization. The resulting strain-induced SRT can be described by calculating the equilibrium orientation of magnetization as a function of strains $u_{m1}$ and $u_{m2}$ via the minimization of $\Delta F$. Since the direction cosines $m_i$ satisfy the relation $m_1^2 + m_2^2 + m_3^2 = 1$, the energy $\Delta F$ appears to be a function of only two variables.

Consider first the case of equal misfit strains ($u_{m1} = u_{m2} = u_m$), which corresponds to a cubic or tetragonal substrate (or thick buffer layer) with the (001)-oriented surface. In order to describe various possible situations, the calculations were performed for cobalt ferrite and iron, where the spontaneous magnetization $\mathbf{M}_s$ is oriented in bulk crystals along one of the edges of the unit cell ($K_{1\sigma} > 0$), and for nickel, where $\mathbf{M}_s$ is directed along the cube diagonal ($K_{1\sigma} < 0$) [19]. In CoFe$_2$O$_4$ films, an abrupt SRT takes place at a critical misfit strain $u_m^*$ defined by the formula

$$u_m^* = \frac{c_{11}}{B_1(c_{11} + 2c_{12})}\left(\frac{1}{2}\mu_0 M_s^2 - \frac{B_1^2}{6c_{11}}\right). \quad (5)$$

With $M_s = 3.5 \times 10^5$ A/m [24], $B_1 = 5.9 \times 10^7$ J/m$^3$ [25], $c_{11} = 2.7 \times 10^{11}$ N/m$^2$, and $c_{12} = 1.6 \times 10^{11}$ N/m$^2$ [26], Eq. (5) gives a small tensile strain of $u_m^* \cong 0.06\%$. In the range of positive misfit strains $u_m > u_m^*$, the magnetization is orthogonal to the film surfaces owing to $B_1 > 0$, whereas at $u_m < u_m^*$ $\mathbf{M}_s$ is parallel to one of the in-plane crystallographic axes, [100] or [010]. This result agrees with the experimental observations of magnetization orientations in CoFe$_2$O$_4$ films of different thicknesses, which were epitaxially grown on MgO [15]. Indeed, the out-of-plane magnetization appears at thicknesses $t \leq 240$ nm ($u_m \geq 0.25\%$), whereas at $t = 400$ nm ($u_m = 0.012\%$) the preferential direction of $\mathbf{M}_s$ lies in the film plane [15]. Hence the prediction of the strain-induced SRT at $u_m^* \cong 0.06\%$ is consistent with the experimental data. The observation of comparable out-of-plane and in-plane remnant magnetizations at $t = 240$ nm may be explained by a distribution of substrate-induced lattice strains in the film.

In iron, the magnetization is much larger than in CoFe$_2$O$_4$ ($M_s = 1.63 \times 10^6$ A/m [27]), whereas the magnetoelastic constant $B_1$ has much smaller absolute value ($B_1 = -3.3 \times 10^6$ J/m$^3$ at $T = 0$ K [28], $c_{11} = 2.42 \times 10^{11}$ N/m$^2$, $c_{12} = 1.465 \times 10^{11}$ N/m$^2$ [29]). As a result, the strain-induced SRT could take place in thick Fe films only at a large compressive strain $u_m^* \approx -23\%$, which is not accessible experimentally (owing to the strain relaxation via generation of misfit dislocations). However, such transition is expected to be possible in ultrathin Fe films with thicknesses close to the critical thickness at which the size-induced SRT occurs [10,13].

In contrast to CoFe$_2$O$_4$ and Fe films, a gradual strain-driven magnetization rotation should take place in Ni films. Indeed, the minimization of the free energy (4) shows that between $u_m^* \cong 0.718\%$ and $u_m^{**} \cong 0.751\%$ the angle between $\mathbf{M}_s$ and the substrate normal gradually changes from 90° to zero [30]. The dependence of the out-of-plane and in-plane magnetization components on the misfit strain is shown in Fig. 1. This behavior is similar to the strain-driven rotation of spontaneous polarization in ferroelectric thin films [31]. It should be noted that in the intermediate strain range $u_m^* < u_m < u_m^{**}$, where $m_1 = m_2 \neq 0$ and $m_3 \neq 0$, the shear lattice strains $u_{13}$ and $u_{23}$ differ from zero so that the phase state of the Ni film is formally monoclinic.

Suppose now that the ferromagnetic film is deposited on a thick ferroelectric substrate with two opposite faces covered by continuous electrodes. For simplicity, we assume that, in the absence of electric field applied to the substrate, the misfit strains $u_{m1}$ and $u_{m2}$ are equal to each other ($u_{m1} = u_{m2} = u_m^0$), which can be achieved by growing the film, if necessary, on an appropriate buffer layer. Owing to the converse piezoelectric effect inherent in ferroelectric materials, the applied electric field **E** creates macroscopic strains $u_{ij} = d_{kij} E_k$ in the substrate having piezoelectric coefficients $d_{kij}$. Via the interfacial coupling in the film/substrate system, the field-induced substrate deformations change the in-plane lattice strains $u_{11}$ and $u_{22}$ in a ferromagnetic film, which may result in the magnetization reorientation. This *electric-field-driven* SRT can be described using Eq. (4) and taking into account variations of the misfit strains $u_{m1}$ and $u_{m2}$ with the field intensity $E$. To maximize the influence of electric field on the film lattice strains, relaxor ferroelectrics Pb(Zn$_{1/3}$Nb$_{2/3}$)O$_3$–PbTiO$_3$ (PZN-PT) or Pb(Mg$_{1/3}$Nb$_{2/3}$)O$_3$–PbTiO$_3$ (PMN-PT) having ultrahigh piezoelectric coefficients [32] should be used as a substrate.

If the electric field is applied along the $x_3$ axis (substrate with the top and bottom electrodes), the misfit strains in a material system may be written in the linear approximation as $u_{m1}(E) = u_m^0 + d_{31}E$ and $u_{m2}(E) = u_m^0 + d_{32}E$. (The matrix notation is employed for piezoelectric coefficients here and below). In the symmetric case ($d_{32} = d_{31}$), the influence of electric field on the magnetization orientation becomes similar to the strain effect discussed above. Therefore, to induce an SRT in the CoFe$_2$O$_4$ film magnetized along the [001] axis ($u_m^0 > u_m^*$), the misfit strain $u_m(E) = u_m^0 + d_{31}E$ should decrease below the critical value $u_m^*$ given by Eq. (5). Since the out-of-plane magnetization changes by $M_s$ at the transition [33], the maximum absolute value of the magnetoelectric susceptibility $\alpha = \mu_0 \Delta M / \Delta E$ equals

$$\alpha_{\max} = \left| \frac{\mu_0 M_s d_{31}}{u_m^* - u_m^0} \right|. \quad (6)$$

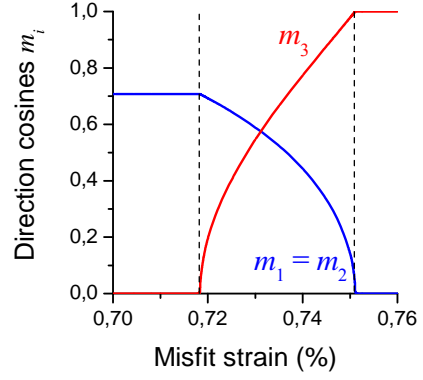

FIG. 1. Direction cosines $m_i$ of the spontaneous magnetization in a thick Ni film as a function of the misfit strain $u_m$ in the heterostructure.

It should be emphasized that the field-induced reduction of the misfit strain may be rather large (~ 1% at $d_{31} \sim -1000$ pm/V [34]). Indeed, it is created by the field directed *along* the substrate polarization, which is limited only by the high dielectric breakdown field $E_b > 100$ kV/cm of PZN-PT or PMN-PT [32], but not by the ferroelectric coercive field $E_c \sim 2$ kV/cm of the substrate.

In contrast to CoFe$_2$O$_4$ and other ferromagnetic films with an abrupt SRT, for which the susceptibility $\alpha$ formally diverges as $u_m^0 \to u_m^*$, the material systems involving Ni films display only a limited magnetoelectric anomaly because here the magnetization rotates gradually in a finite strain range. To evaluate the upper bound of $\alpha$, we studied the electric-field-induced magnetization reorientation in Ni films for a special heterostructure, where the initial strain $u_m^0$ is equal to the critical strain $u_m^{**} \cong 0.751\%$ corresponding to the right boundary of SRT in Fig. 1. Using the dependence $m_3(u_m)$ plotted in Fig. 1 and the relation $E = (u_m - u_m^0)/d_{31}$ between the electric field and the misfit strain, we calculated the magnetoelectric susceptibility as $\alpha(E) = \mu_0 M_s [m_3(E) - 1]/E$. The results of calculations (see Fig. 2) demonstrate two remarkable features: (i) the susceptibility reaches very high level of $10^{-6}$ s/m already at a small field $E \approx 20$ V/cm, and (ii) the magnitude of $\alpha$ weakly depends on the field intensity up to $E \sim 5$ kV/cm. The maximum theoretical value $\alpha_{max} = 1.86 \times 10^{-6}$ s/m is almost 10 times larger than the magnetoelectric susceptibility $\alpha \sim 2.3 \times 10^{-7}$ s/m achieved recently in the $La_{0.67}Sr_{0.33}MnO_3/BaTiO_3$ heterostructure [7]. This value also greatly exceeds the susceptibility reported for the $La_{0.7}Sr_{0.3}MnO_3$ film grown on PMN-PT [6] and is several orders of magnitude larger than the susceptibilities of magnetoelectric crystals [1].

The electric field can be applied to the substrate in a direction parallel to the film surfaces as well. In this case the strains $u_{m1}(E)$ and $u_{m2}(E)$ become very different so that the magnetoelectric effect acquires new features. If the field is directed along the [100] crystallographic axis of the film, the misfit strains can be evaluated as $u_{m1}(E) = u_m^0 + d_{33}^* E$ and $u_{m2}(E) = u_m^0 + d_{31}^* E$, where $d_{in}^*$ are the substrate piezoelectric coefficients defined in the reference frame $(x_1^*, x_2^*, x_3^*)$ with the $x_3^*$ axis oriented along the field direction and the $x_1^*$ axis parallel to the interface. For $CoFe_2O_4$ films, the calculation shows that the SRT can be induced at $u_m^0 > u_m^*$ by an in-plane electric field directed along the substrate polarization. The critical field intensity $E_c$ is given by the relation

$$E_c = -\frac{(c_{11} + 2c_{12})(u_m^0 - u_m^*)}{[(c_{11} + c_{12})d_{31}^* + c_{12}d_{33}^*]}, \qquad (7)$$

where the denominator is negative since $d_{31}^* < 0$ and its magnitude is close to $d_{33}^*/2$ [34]. When the applied field exceeds $E_c$, the magnetization flips onto the film plane and becomes parallel to the $x_2$ axis. The maximum magnetoelectric susceptibility $\alpha_{max} = \mu_0 M_s / E_c$ is proportional to $(u_m^0 - u_m^*)^{-1}$ and exceeds $10^{-6}$ s/m at $u_m^0 - u_m^* < 0.056\%$ in the case of $CoFe_2O_4$ film grown on PZN-4.5%PT ($d_{33}^* = 2000$ pm/V, $d_{31}^* = -1000$ pm/V [34]).

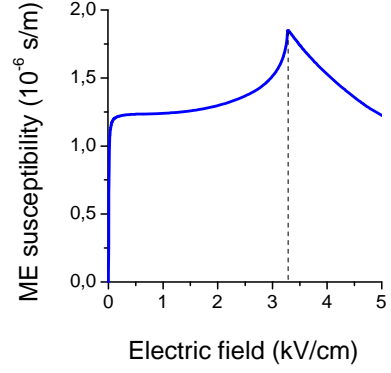

FIG. 2. Magnetoelectric susceptibility of the Ni/PZN-4.5%PT heterostructure as a function of electric field applied to the substrate in the direction orthogonal to film surfaces. The dashed line corresponds to the end of SRT. The initial misfit strain is taken as $u_m^0 = 0.7511\%$, and the substrate piezoelectric coefficient $d_{31} = -1000$ pm/V.

For Ni films, the minimization of the energy density (4) demonstrates that a gradual magnetization rotation takes place in these films when an in-plane electric field is applied to the ferroelectric substrate. Figure 3 shows the direction cosines of magnetization as a function of the field intensity for the special heterostructure introduced above ($u_m^0 = u_m^{**}$). Remarkably, the magnetization rotates in the (100) crystallographic plane ($m_1 = 0$), but not in the $(1\bar{1}0)$ one ($m_1 = m_2$), as it happens under the influence of electric field orthogonal to the film/substrate interface. The variation of the magnetoelectric susceptibility $\alpha(E) = \mu_0 M_s [m_3(E) - 1]/E$ with the field intensity is similar to the dependence shown in Fig. 2. However, the maximum theoretical susceptibility $\alpha_{max} = 0.285 \times 10^{-6}$ s/m reached at the end of SRT is considerably smaller than in the case of electric field orthogonal to the interface.

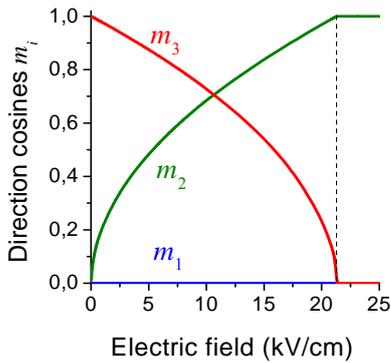

FIG. 3. Direction cosines $m_i$ of the spontaneous magnetization in a thick Ni film as a function of electric field applied to the PZN-4.5%PT substrate along the $x_1$ axis parallel to the interface. The initial misfit strain in the heterostructure is assumed to be $u_m^0 = 0.7511\%$.

Thus, the spin-reorientation transitions may be induced in epitaxial ferromagnetic films by a moderate electric field applied to a ferroelectric substrate. The resulting magnetoelectric effect, which is mediated by the mechanical film/substrate interaction, increases dramatically when the misfit strain in the heterostructure becomes close to a critical value corresponding to the strain-induced SRT. The magnetoelectric susceptibility of such ferromagnetic/ferroelectric heterostructures may reach giant values exceeding $10^{-6}$ s/m.